# Viewing heat through ice: an infrared camera monitors hydrogel freezing and thawing during cryoapplication

Gennadiy O. Kovalov[a,*], Mykola O. Chyzh[a], Vyacheslav Yu. Globa[a], Oleksandr F. Todrin[a], Galyna V. Shustakova[b], Eduard Yu. Gordiyenko[b], Yuliya V. Fomenko[b], Oleh V. Ivakhnenko[b], Polina O. Kofman[b], Sergey N. Shevchenko[b]

[a] *Institute for Problems of Cryobiology and Cryomedicine of the National Academy of Sciences of Ukraine, Postal address: 23, Pereyaslavska str., Kharkiv, 61016, Ukraine*

[b] *B. Verkin Institute for Low Temperature Physics and Engineering of the National Academy of Sciences of Ukraine, Postal address: 47 Nauky Ave., Kharkiv, 61103, Ukraine*

**Abstract**

Cryosurgery employs a safe and relatively simple technique of exposure and is an advantageous and highly rated method. For its effective application, it is necessary to control both the volume of the expanding freezing zone and volumetric thermal field dynamics. The aim of this study was to perform a thermal imaging study of freezing and thawing in a model system (gel phantom) to predict the dynamics of the freezing zone during cryodestruction of biological tissues *in vivo*. Here, the thermal imager is an effective tool for demonstrating the surface temperature distribution. We have studied how the observed infrared image relates to the distribution and change of the thermal field in depth. For this purpose, we created test measuring equipment for simultaneous analysis of the dynamics of thermal fields on the surface, video recording of freezing and thawing on the surface as well as in the depth of the gel phantom, measuring the temperature at any given point in the depth and modeling in the zone of low-temperature exposure of vessels with different blood flow parameters. It was revealed that with a modeled vessel in the low-temperature exposure zone, the surface thermal fields deformed and they gained the shape of butterfly wings. Our experimental study in a gel phantom is supported by numerical calculations, demonstrating how the freezing zone and thermal isotherms on the surface and in depth evolve under real conditions, thereby providing a basis for assessing the cryoeffect time and intensity in practice.

# 1. Introduction

Cryosurgery is a minimally invasive, effective and safe method that is an alternative to invasive surgery. Cryosurgery is intended for targeted and controlled destruction of affected tissues by applying cold and is indicated for the treatment of benign, precancerous and malignant neoplasms of skin and soft tissues [6, 21]. Advantages of cryosurgery: a fairly safe and relatively simple intervention technique is used; excellent cosmetic and therapeutic results; low complication rate; short treatment; few contraindications; low cost; high healing rates even in «difficult» areas; can be performed on an outpatient basis; repeated as needed; can be used in elderly, inoperable patients; combined with other methods of treating oncology diseases [24].

For successful cryosurgery, it is necessary to control the freezing zone volume and temperature during low-temperature exposure, take into account the factors affecting tissue freezing and thawing: cooling rate, final tissue temperature, cryoexposure time, thawing rate depending on the type of tissue, etc. [15]. When performing cryosurgery on affected areas of skin and soft tissues, the volume of frozen tissue can be controlled visually or by palpation, but these methods are not reliable and predictable. There are tools for monitoring low-temperature exposure (frozen tissue volume and temperature). These are direct and remote thermometries, ultrasound diagnostics, computed tomography, magnetic resonance imaging, electrical impedance tomography, optical coherence tomography, etc. [15]. Each method has its own advantages and disadvantages. Infrared thermography (IRT) can be considered as a promising non-invasive control of the temperature field in skin cryosurgery [11]. Temperature fields on the surface provide information on temperature distribution in the depth of the object under study [17], therefore, thermal imaging of the object surface can be considered as encouraging way for predicting the dynamics of the freezing zone parameters in the depths of biological tissues during their cryodestruction.

In order to cause complete cell death during cryoablation, it is necessary to achieve a temperature below –20°C. Full consensus on the in vivo temperatures required for complete cell death in some organs has not yet been achieved; lethal temperatures for different tissues range from –20°C to –40°C. Porgel M. et al. conducted experiments on rats using a three-fold freeze-thaw cycle. Histological studies showed that cells that were within the –20°C isotherm died completely [16]. Torre D. [20] also reports that a temperature of –25°C is necessary for skin devitalization.

The presence of blood vessels does not prevent the spread of the ice front in tissues. Ice crystal growth in microvessels with an arteriole diameter (10–15 μm, 4–35 μm capillaries) occurs differently than in large vessels [3]. The difficulties in describing heat transfer in living tissues are associated with the need to take into account heat transfer by blood and are determined by both the complexity of the blood vessel system [2] as well as the change in perfusion intensity depending on the degree of tissue damage and the state of the body.

Progress in cryosurgery is impossible without experimental studies that allow a better understanding of the processes occurring during freezing and thawing of biological tissues. Assuming the serious ethical problems that arise when performing research in animals, *in vitro* experiments are preferable, the data from which could be extrapolated to biological tissues *in vivo*. The use of a gel phantom is promising for modeling various freezing and thawing parameters during cryodestruction of skin. A gel phantom made from a 5% aqueous solution of gelatin has good transparency, stable thermodynamic properties and is close in parameters to biological tissues. Thus, the use of a gel phantom as a model system allows studying the temperature parameters on its surface and in depth simultaneously with visual observation of the process of changing the ice zone size.

The purpose of this study was a thermal imaging study of freezing-thawing processes on the surface of a gel phantom to predict the dynamics of the freezing zone during cryodestruction of biological tissues *in vivo*.

**2. Materials and methods**

**2.1. Object of study**

These were *in vitro* experiments, performed in a model system consisting of 5% gelatin and 95% water (hereinafter referred to as hydrogel) with a temperature of (20 ± 2) °C prior to cryoeffect onset. To obtain the hydrogel, gelatin granules (5% of the mass) were soaked in distilled water at room temperature for 2 hrs. Then the mixture was heated to 60 °C to obtain a homogeneous solution [22], afterwards the solution was kept at room temperature for 20-22 hrs until complete swelling.

**2.2. Measuring setup**

A special multichannel measuring setup was created for the studies, allowing for modeling and analyzing the *in vitro* freezing and thawing. The setup allows simultaneously and in real time to: (i) perform cryoeffect in the model system using various cryoinstruments (those of application, spray and penetrating effects), (ii) observe and record thermal images of the surface of the model systems using a thermal imager, (iii) video record the freezing-thawing processes in transparent model systems from various directions using video cameras; (iv) measure temperature dynamics at selected points in the model system volume. It also can vary the composition and temperature of the model system to simulate various biological objects, place vessel models with different blood flow parameters in the low-temperature exposure zone, etc. The diagram and appearance of the setup are shown in Fig. 1.

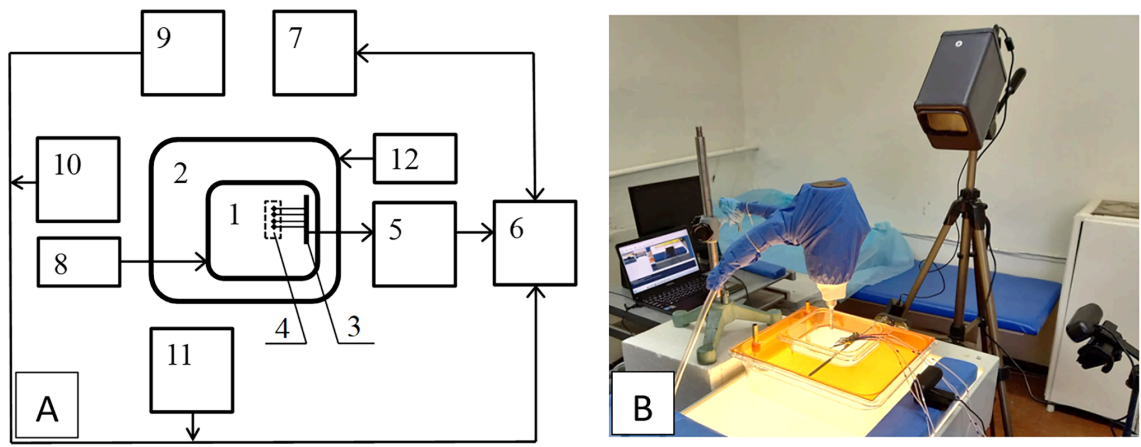

Fig. 1. Scheme (A) and external appearance (B) of the measuring setup for modeling cryoeffects *in vitro*. (A): 1 – container with model liquid, 2 – container with coolant, 3 – device for fixing thermometers, 4 – set of resistance thermometers, 5 – multichannel analog-to-digital signal converter, 6 – computer for visualization and automatic data recording, 7 – thermal imager, 8 – cryoinstrument, 9, 10, 11 – video cameras, 12 – circulation thermostat with pump.

### 2.3. Experiment design and procedure

Design and course of the experiments to study the freezing and warming processes during *in vitro* cryodestruction were as follows (Fig. 1A): hydrogel with a given temperature was placed in a container (1); container (1) was immersed in a container (2) filled with liquid with the same temperature. The temperature of the liquid in the container (2) was maintained using a circulation thermostat with a pump (12), which constantly pumped the liquid through the container (2). Thermometers (4) were placed using a fixing device (3) at given points of the hydrogel and connected to a multichannel analog-to-digital signal converter (5). The thermal imager (7) was fixed with a tripod at a given angle to the hydrogel surface and distance to it. Video cameras (9), (10), (11) were placed in positions that provided the required directions of viewing the cooling zone. A multichannel analog-to-digital signals converter (5), a thermal imager (7) and video cameras (9), (10), (11) were connected to a personal computer (6). Cryoinstrument was fixed on a tripod with the possibility of smooth height adjustment (8). After testing the operation of the devices and recording the initial data, liquid nitrogen was poured into the cryoinstrument. After cooling the working surface of the applicator, low-temperature action on the surface of the hydrogel began. Simultaneously, automatic thermographic shooting with a thermal imager (7), direct thermometry with thermometers (4) and video filming of the freeze-thawing with video cameras (9), (10), (11) were carried out. The duration of low-temperature impact on the surface of the hydrogel was 30 minutes, followed by natural thawing for 30 minutes.

### 2.4. Cryoinstrument

To accomplish cryoeffect, we used an original tool (manufactured at the Institute for Problems of Cryobiology and Cryomedicine of the National Academy of Sciences of Ukraine) with a flat copper applicator with a diameter of 8.0 mm, cooled with liquid nitrogen. The cryoinstrument's entry into the operating mode before the start of low-temperature exposure was controlled by measuring the temperature of the applicator working surface (192.0 ... -194.8) °C.

### 2.5. Direct thermometry

The temperature dynamics at selected points in the hydrogel volume were studied using platinum thermal resistance sensors (OWEN DTS-100P), measuring temperature within the range of (-196 ... +250) °C with an accuracy of ±0.3 °C. The sensors were connected to the Owen MV-110 analog input module, the data from which were transmitted to the Owen AC4 interface converter, and then to the PC. Data recording and analysis were performed using Owen Process Manager 1.2 and Owen Report Viewer 1.2 software. Thermometers were fixed in the hydrogel at a depth of (10-15) mm at different distances from the point of contact of the cryoapplicator center with the surface: resistance thermometer #1 – at a distance of approximately 30 mm, resistance thermometer #2 – at a distance of 40 mm, resistance thermometer #3 – at a distance of 50 mm. We estimate the accuracy of thermometer localization as ± 2 mm. Figure 2 shows an image (side view) of the inner container with hydrogel at the 10th minute of cooling. The formed ice lump with a shape close to hemispherical and three thermometers located at different distances from the frozen zone can be seen.

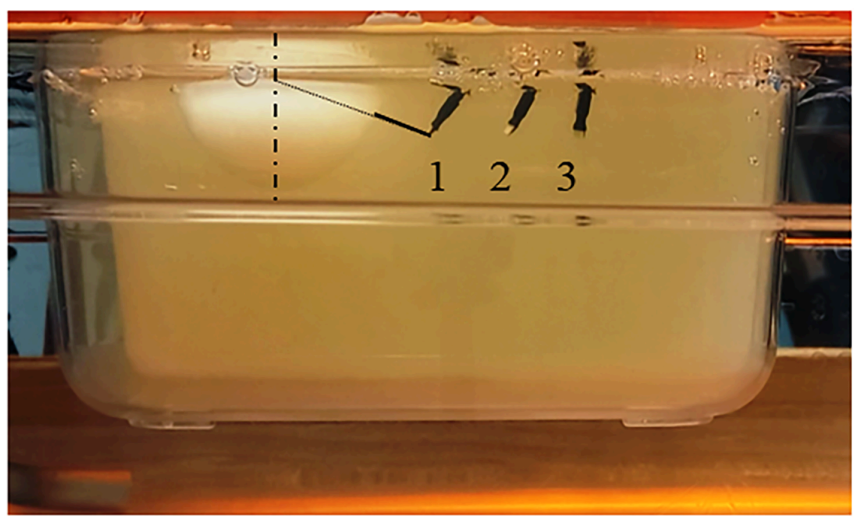

Fig. 2. Side view of the inner container with hydrogel during cryoexposure. The resultant lump of ice with a shape close to hemispherical and three thermometers located at different distances from the frozen zone can be seen.

### 2.6. IR and video cameras

The passive infrared thermography was used to analyze the dynamics of thermal fields on the hydrogel surface. We exploited an analyzer of low-temperature (down to -150 °C) thermal fields, i.e. a thermal imager based on a single-element semiconductor detector cooled to the temperature of liquid nitrogen. The device was developed at the B.I. Verkin Institute for Low Temperature Physics and Engineering of the National Academy of Sciences of Ukraine [5]. The thermal imager temperature sensitivity was of 0.1 °C within the spectral range of 7…14 μm, a frame rate of 0.5 Hz, a field of view (22H × 22 V) deg, spatial resolution of 1.5 milliradians.

The device records absolute surface temperature values with an error of: $\delta T \approx \pm 1$ °C (in the range of measured temperatures T = (+50…-30) °C), $\delta T \approx \pm 1.5$ °C (at a temperature level of T = -50 °C) and $\delta T \approx \pm 4.5$ °C (at a temperature level of T = -150 °C). To eliminate infrared reflection from the hydrogel surface of high-temperature objects, the thermal imager was fixed with a tripod at an angle of 70 degrees to the hydrogel surface at a distance of 1 m. This angle and distance remained unchanged in all experiments. The software specially developed for this task had many functions useful for the task at hand, including the ability to continuously record a thermographic film in digital format for half an hour at a frequency of 0.5 frames per second. The thermal image is output to the PC screen via the USB interface.

Before the start of the research, the thermal imager was tested with the Fluke Portable Infrared Calibrator 9133 metrology stand (Fluke Corporation, USA) in the range of measured temperatures T = (+50…-30) °C. For the temperature range (-30…-150) °C, the system was calibrated using the original blackbody model (BBM) cooled with liquid nitrogen. The emissivity of the BBM surface: $\mathcal{E}=0.98$. The BBM surface temperature was controlled by a contact thermometer. For thermal imaging temperature measurements, the emissivity of the liquid and frozen hydrogel surfaces was $\mathcal{E}=0.95$ and $\mathcal{E}=0.96$, respectively. These values were obtained by comparing the thermal images of the hydrogel and BBM surfaces at the same temperatures (+20 and -30) °C. Similar values have been reported [14].

For video recording of the dynamics of ice growth and melting in the hydrogel, three ATech PK 935HL cameras with a resolution of Full HD 1920 x 1080 (ATech, China) were used, providing different viewing directions.

### 2.7. Blood flow modeling

A polyvinyl chloride vessel model with an outer diameter of 2.7 mm (Nelaton male urological catheter size 8 (Medicare Great Britain) was used to model the blood flow in the low-temperature exposure zone. The vessel model was placed in the hydrogel to a depth of (15±2) mm (Fig. 3). The distance between the entry and exit points of the vessel from the hydrogel was approximately 80 mm. Turusol solution (Yuria-Pharm LLC, Ukraine) was used as an irrigation fluid. The fluid flow rate in the modeled vessel was approximately 2.8 ml/min, the inlet temperature was (20 ± 2) °C.

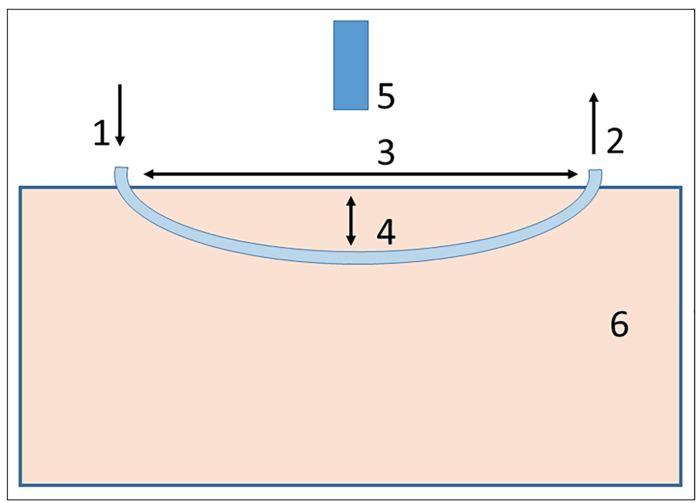

Fig. 3. Schematic representation of the arrangement of the vessel model in the hydrogel: 1, 2 – directions of fluid movement; 3 – distance between the entry and exit points of the vessel model into the hydrogel; 4 – distance from the hydrogel surface at the point of cryoeffect to the vessel model; 5 – cryoapplicator; 6 – hydrogel.

### 2.8. Data processing

The obtained numerical data were processed by statistical analysis using the software Microsoft Excel 2010 (Microsoft, USA) and Statistica 10.0 (StatSoft, USA) and were presented as M ± S, where M is the sample mean value of the parameter, S is the standard deviation for five experiments (n = 5). The Mann-Whitney criterion was used to compare the parameters. The reliability of differences between the average values in the groups was accepted at a significance level of $p < 0.05$.

The standard deviation of the values from the mean is presented in all experimental dependencies.

### 3. Results and discussion

### 3.1. Freezing of hydrogel under low-temperature exposure

The results of the experiments showed that from the first seconds of low-temperature exposure, a volumetric ice lump was formed in the hydrogel, the base of which on the surface shaped as a round ice spot, that corresponds to the reported data [8]. To analyze the dynamics of temperature distribution on the hydrogel surface during cryoexposure, the obtained thermographic films were used. It was found that all isotherms had the shape of concentric circles. These results coincide with our own data obtained earlier *in vivo* [10].

The dynamics of changes in the diameters of isotherms (0, –20, –40) °C on the hydrogel surface were measured. These isotherms were chosen due to their great importance for skin cryosurgery. The 0 °C isotherm is not associated with the release of phase transition heat, since gelatin hydrogel (like biological tissues) freezes in the temperature range below 0 °C. However, this isotherm may be of interest as an extreme point before the onset of freezing of both the hydrogel and biological objects. With certain assumptions, the 0°C isotherm can be considered the boundary of the frozen zone on the surface *in vivo* [9], and the –20°C and –40°C isotherms provide an idea of the distribution of lethal temperatures for various biological tissues [18, 19].

Figure 4 shows a thermal image of the hydrogel surface at the 25th minute of cryoexposure. The shape and current locations of the isotherms are schematically shown by dashed lines: white line - isotherm (-40 °C), yellow (-20 °C), red (0 °C).

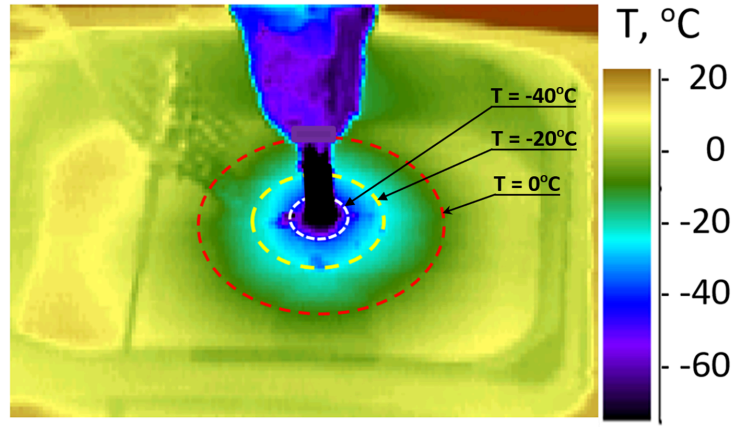

Fig. 4. Distribution of thermal fields on the hydrogel surface at the 25th minute of cryoexposure. The red, yellow and white dashed lines schematically show the shape and current location of the isotherms (0, -20, -40) °C, respectively.

Figure 5 shows the increase in the diameters of the surface isotherms (0, –20, –40) °C during 30-minute cooling, - the red, green and blue curves, respectively. The corresponding approximants are also shown in the figure by dashed lines. Figure 5 evidently demonstrates that, in general, all the curves have a similar appearance, and the best approximant for them is the logarithmic dependence:

$$d(t) = A \ln t + B, \quad (1)$$

where d is the isotherm diameter (mm), t is time (min), A and B are coefficients.

Differentiating equation (1) with respect to time, we obtain the dependence of the rate of increase of the diameters $V$ (mm/min) for these isotherms at each moment of time:

$$V = A / t. \quad (2)$$

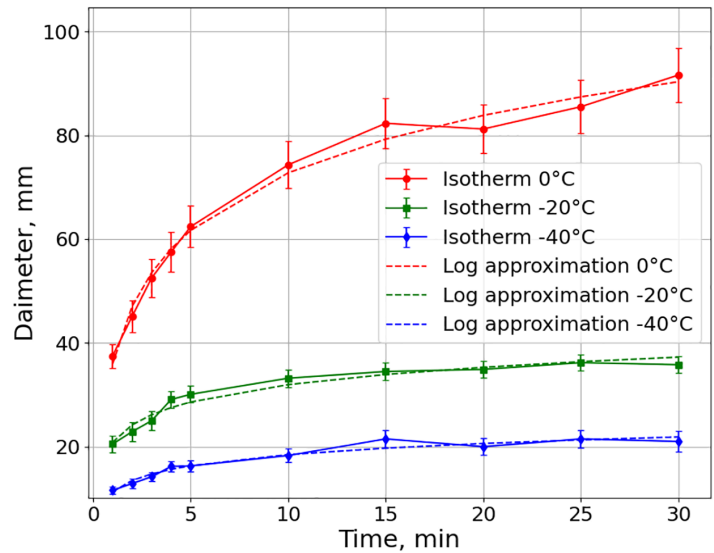

Fig. 5. Growth of isotherm diameters (0, –20 and –40) °C under low-temperature action on the hydrogel surface by a cryoapplicator. Dashed lines are approximants. Mean (± S) of the diameters measured in five experiments.

The values of the coefficients of equations (1) and (2) for three isotherms are presented in the Table.

Table. Coefficients of equations 1 and 2.

| Coefficients | A (mm) | B (mm) |
| --- | --- | --- |
| Isotherm 0°C | 15.97 | 35.99 |
| Isotherm –20°C | 4.82 | 20.86 |
| Isotherm –40°C | 3.10 | 11.34 |

The dependences of the rates of increase in the diameters of the surface isotherms (0, –20 and –40) °C (red, green and blue curves, respectively), calculated using equation (2), are shown in Fig. 6. It is evident that the rate of increase in the diameters of all isotherms decreases sharply during the first 5 minutes of cryogenic exposure, later it asymptotically tends to zero.

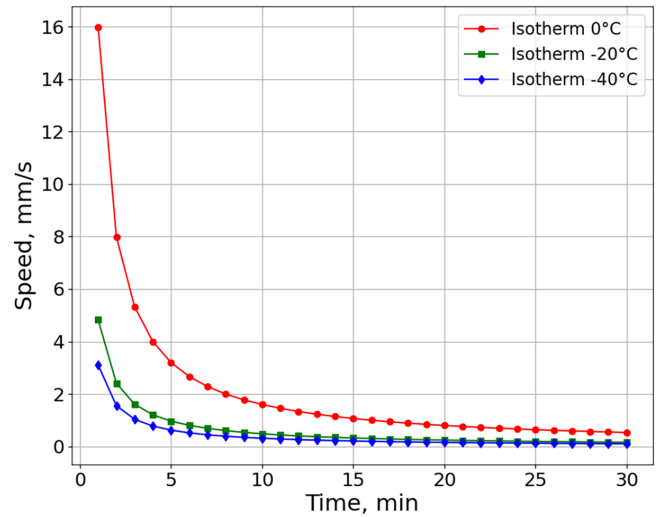

Fig. 6. Dependence of the rates of increase of the diameters for isotherms (0, –20 and –40)°C on the hydrogel surface during cryoexposure (red, green and blue curves, respectively).

According to direct thermometry data (thermometers #1, 2 and 3, - see Fig. 2), it was established that at any moment of cryoexposure at a depth of 10-15 mm of the hydrogel, a temperature difference is observed between points with various localizations (Fig. 7). However, it is possible that heat inflows through the wires of the thermometers, which can be neglected at the beginning of cryoexposure, can affect the values of the measured temperature at deep points of the hydrogel as it cools.

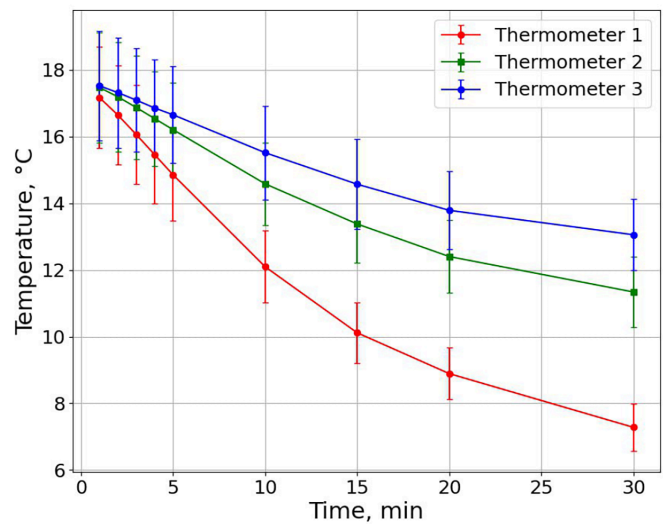

Fig. 7. Temperature dynamics at three points at a depth of (10-15) mm from the hydrogel surface during 30 min of exposure. Mean (± S) of the temperatures measured in five experiments.

The lowest temperature, as expected, was observed in the zone closest to the edge of the cryoapplicator (thermometer #1, red curve), the highest was done in the zone farthest from its edge (thermometer #3, blue curve), and the difference in thermometer readings increased with a rise in cryoexposure time, which is consistent with the data [17].

Let us consider «ideal» conditions, when the hydrogel volume is very large, the cooling capacity of the cryoinstrument is constant and there are no additional factors. After a certain period of time, equilibrium will arise between the amount of cold from the cryoapplicator and heat from the hydrogel. Therefore, the growth of the ice lump should first slow down over time and eventually stop. At the same time, equilibrium will also arise for the temperature distribution in the hydrogel. Under these conditions, the dynamics of the parameters of the surface and volumetric thermal fields coincide.

In our experiments, the conditions were not ideal. The side surface of the cryoapplicator was an additional source of cold due to its insufficient thermal insulation. We used a limited volume of hydrogel, to which heat came from the environment owing to convection, radiation and through the wires of the thermometers.

Additional factors of the «real» experiment to varying degrees influenced the thermodynamic processes on the surface and in the depth of the hydrogel. Thermography (Fig. 5) and direct thermometry data (Fig. 7) show the emergence of differences in the parameters of surface and volumetric thermal fields during cryoeffect. However, the influence of additional factors with a cryoeffect duration of no longer than 3 minutes is insignificant and can be neglected at this cooling intensity. This statement of ours is based on a sharp increase in the diameters of isotherms on the hydrogel surface and the rapid growth of the ice lump during this period of time.

Therefore, we believe that when planning experiments wherein the freezing zone in the depth of an object will be estimated based on the temperature fields on its surface, one should strive to select a short duration of cryoexposure.

We do not know the temperature of the growing ice lump edge. Only the temperature range of melting of frozen hydrogel is known (-0.1…-14) °C [1]. Moreover, the hydrogel supercooling at the beginning of cryoexposure, when the cooling rate was maximum one, cannot be ruled out. Thus, the observed edge of the growing ice lump could have a temperature below the freezing onset temperature, and this temperature changed during cryoexposure. Due to methodological difficulties, we have not yet been able to accurately measure the temperature of the ice lump edge. In further studies, we hope to overcome these difficulties and measure the temperature of the ice lump edge at different stages of cryoexposure and thawing.

## 3.2. Freezing of hydrogel with a blood vessel model

In the presence of a model vessel with blood flow in the cryoeffect zone, the dynamics of ice lump formation in the depth of the hydrogel generally corresponded to the one described without modeling the vessel. According to visual observation, the ice lump also grew throughout the low-temperature exposure in the presence of a vessel. The increase in the size of the ice lump in the volume of the hydrogel was estimated by the decrease in the distance from the visible edge of the ice lump to resistance thermometer #1. For this purpose, a straight line was drawn connecting the center of contact of the cryoapplicator with the surface and the thermometer, and the length of the segment of this straight line between the visible edge of the ice surface and the thermometer was measured (see Fig. 2). The dynamics of the decrease in this segment is shown in Fig. 8. At the same time, no significant differences were found between the values in the presence of the vessel model (red curve) and its absence (green curve).

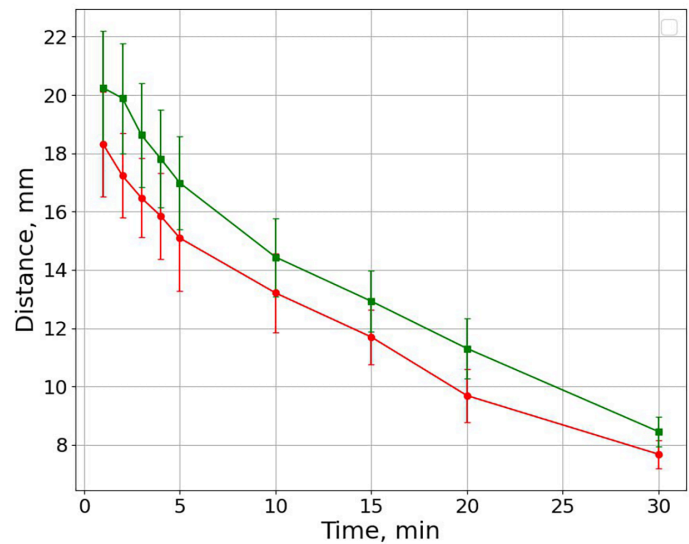

Fig. 8. Dynamics of the decrease in the distance from the ice lump edge to thermometer #1 under low-temperature on the hydrogel surface in the presence of a vessel with blood flow (green curve) and without it (red curve). Mean (± S) of the distances measured in five experiments.

However, the shape of the ice spot on the hydrogel surface differed notably from that «without a vessel». In the presence of a model vessel with blood flow, the ice spot quickly lost its rounded shape and acquired the appearance of butterfly wings, which is due to the heating of the surface near the entry and exit points of the vessel into the hydrogel. It should be noted that the deformation of all isotherms started from the vessel model incoming edge. This can be explained by a higher surface temperature of the tube with liquid at the inlet than at the outlet from the hydrogel due to its cooling in the cryogenic effect zone. Figure 9 (A, B) illustrates thermograms of a fragment of the hydrogel surface, showing the temperature fields on the surface in the presence of a model of a vessel with blood flow after 2 min (A) and after 15 min (B) of low-temperature exposure. The red and brown rings mark the entry and exit of the vessel into the hydrogel, respectively. The current shape and locations of the isotherms are shown by dashed lines: isotherm -40 °C (white line), isotherm -20 °C (yellow line), isotherm 0°C (red line).

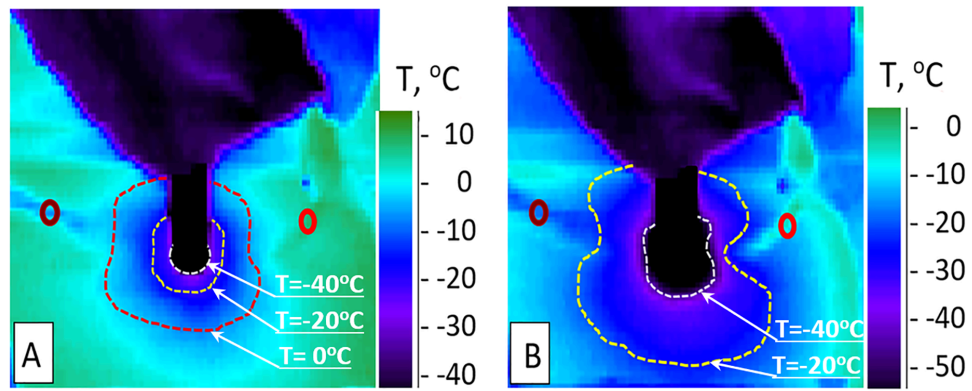

Fig. 9. Distribution of temperature fields on the hydrogel surface in the presence of a model vessel with blood flow after 2 min (A) and after 15 min (B) of low-temperature exposure. The red and brown rings mark the vessel's entrance and exit into the hydrogel, respectively. The instantaneous shape and locations of the isotherms are shown by dashed lines: -40 °C isotherm (white line), -20 °C isotherm (yellow line), 0 °C isotherm (red line).

The surface isotherms of 0 °C and –20 °C were deformed from the first seconds of low-temperature exposure. The surface isotherm of –40 °C had a rounded shape at the beginning of cryoeffect, but after 2 min of cryoexposure it also acquired the appearance of butterfly wings.

A comparison of the diameters of the surface isotherms of –40°C obtained with and without the vessel during the first 2 min of low-temperature exposure did not reveal any strong differences between them. Figure 10 shows the growth of the diameters of these isotherms with a similar character. The findings indicate the absence of a significant effect of heat supply from the vessel on the formation of the –40°C isotherm only during short-term low-temperature exposure.

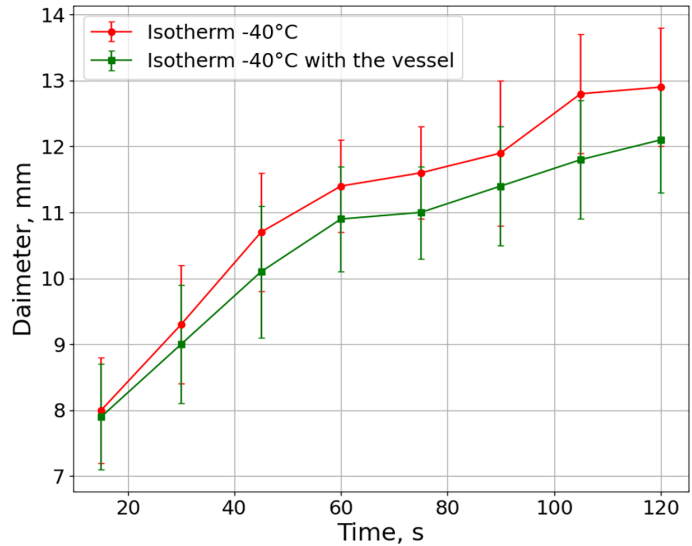

Fig. 10. Dynamics of the diameters of the surface isotherms at –40°C under low-temperature exposure with a model vessel with blood flow (red curve) and without it (green curve). Mean (± S) of the diameters measured in five experiments.

Our findings can be logically explained by the heating of the hydrogel by the influx of heat through the vessel walls from the flowing liquid. This heat is sufficient to change the shape of the isotherms of the 0 °C and –20 °C surface already at the beginning of the cryoeffect, and after 2 minutes of cryoexposure, the –40 °C isotherm as well. Consequently, the cooling capacity of the cryoinstrument was insufficient to compensate for the additional influx of heat and achieve uniform freezing of the hydrogel near the vessel model to critical temperatures in cryosurgery. The fact that during low-temperature exposure the shape of the thermal field limited by the isotherm of -40°C remained rounded only for 2 minutes makes questionable the correctness of extrapolating the characteristics of the temperature fields on the surface to those of the temperature fields in the depth of the hydrogel during exposure for more than 2 minutes (under our thermodynamic conditions, characteristics and location of the vessel model, etc.).

The obtained data demonstrate the importance of taking into account the presence of vessels in the low-temperature exposure zone and their topographic and anatomical features during thermographic monitoring during cryosurgical intervention. Failure to comply with this requirement may lead to an erroneous assessment of the lethal isotherm depth and shape, which is based on estimations of thermal field on the surface. This mismatch will affect premature termination of low-temperature exposure, which threatens poor cryodestruction of the pathological formation.

### 3.3. Natural warming with and without a blood vessel model

A visual assessment of the dynamics of the increase in the distance from the ice lump edge to the resistance thermometer #1 during natural warming indicates a slow decrease in the ice lump size, and the thermometer readings indicate an increase in temperature at this point, regardless of the presence or absence of a vessel model. The minimum rate of temperature increase was observed at the point closest to the freezing zone (data from thermometer #1).

Analysis of the temperature fields on the hydrogel surface during thawing of the freezing zone without a vessel shows that the isotherms had the shape of concentric circles; the same was observed when low-temperature exposure was applied. Both during thawing and freezing, at each moment in time the temperature was lower, the closer the point was to the center of the freezing zone. In the thermogram (Fig. 11), the current position of the isotherms of -20 °C and 0 °C, respectively, during natural warming after 30 minutes of low-temperature exposure, is schematically marked in yellow and red.

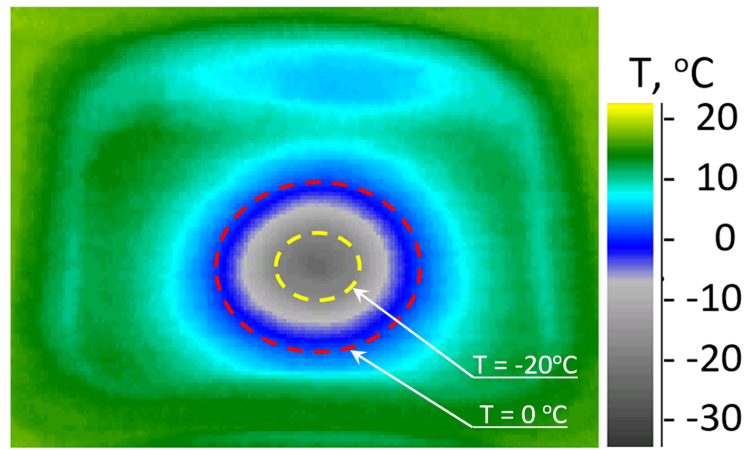

Fig. 11. Distribution of thermal fields on the hydrogel surface during natural thawing after low-temperature exposure for 30 min. Yellow and red dashed lines schematically show the concentric shapes and locations of isotherms (-20 °C) and 0 °C, respectively.

The diameter of the surface isotherm -20 °C was found to decrease to the greatest extent in the first minute after the end of low-temperature exposure. Figure 12 shows the dependence of the decrease in the diameter of the surface isotherm -20°C over 30 minutes of natural warming.

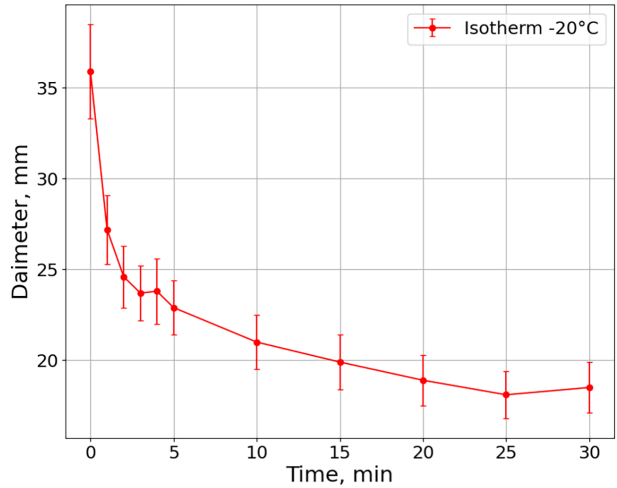

Fig. 12. Dynamics of the diameter of the surface isotherm (–20°C) during natural thawing of the freezing zone after low-temperature exposure. Mean (± S) of the diameters measured in five experiments.

The highest thawing rate was observed in the first seconds after the end of the low-temperature exposure. The dynamics of the change in the diameter of the surface isotherm (–40°C) during natural thawing of the freezing zone could be traced only for 22 s. Figure 13 shows the dependences of the decrease in the diameters of the surface isotherms of –20 °C and –40°C during the first minute of natural thawing of the freezing zone after the low-temperature exposure end.

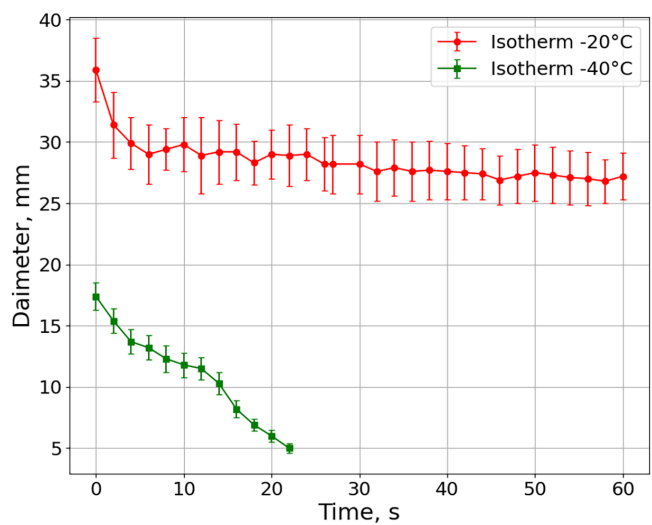

Fig. 13. Dynamics of the diameters of the surface isotherms of –20 and –40°C (red and green curves, respectively) during the first minute of natural thawing of the freezing zone after low-temperature exposure. Mean (± S) of the diameters measured in five experiments.

The data presented above correspond well with the dynamics of changes in the minimum surface temperature of the freezing zone, shown in Figs. 14 and 15. As Fig. 14 shows the sharpest changes in the minimum surface temperature occurred in the first minute after the low-temperature exposure cessation. During this time, the temperature increased from $T_{min} = (-110.0 \pm 10.9)$ °C to $T_{min} = (-33.0 \pm 3.2)$ °C.

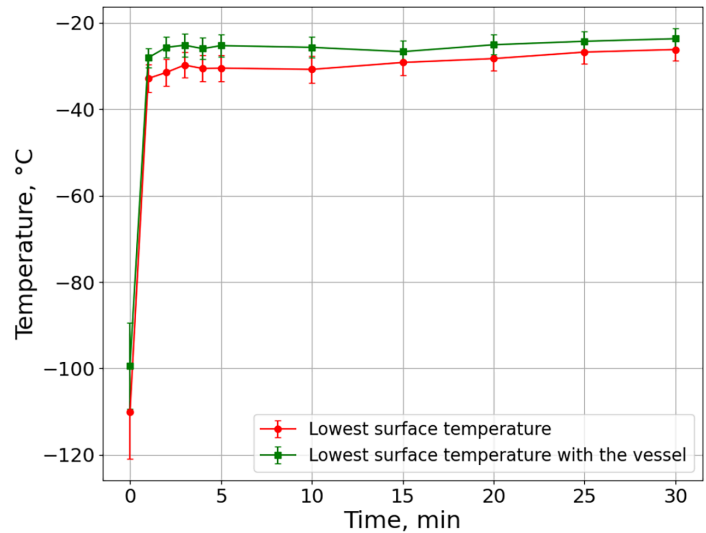

Fig. 14. Dynamics of the minimum surface temperature of the freezing zone during its natural thawing with a vessel (red curve) and without it (green curve) in the freezing zone. Mean (± S) of the temperatures measured in five experiments.

The results of the analysis of the dynamics of the lowest temperature over 1 min allow the conclusion about its greatest changes occurred in the first seconds of warming both with a vessel and without it (Fig. 15).

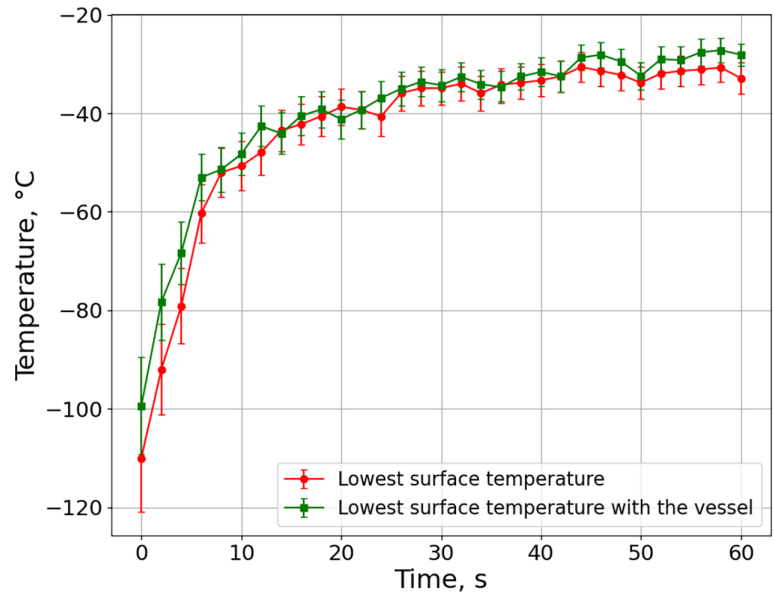

Fig. 15. Dynamics of the lowest temperature of the freezing zone surface during the first minute of its natural thawing with a vessel (red curve) and without it (green curve) in the freezing zone. Mean (± S) of the temperatures measured in five experiments.

The sharp rise in the hydrogel surface temperature at the first moments after the cessation of cryoeffect (Fig. 11-15) is apparently caused by the cooling source removal. According to the results of statistical analysis, no significant differences were found between the temperature values at the corresponding observation times with or without a model vessel during the entire thawing period.

It should be noted that when modeling a vessel with blood flow, the area with the lowest surface temperature did not coincide with the freezing zone geometric center (which corresponded to the center of the cryoapplicator), as was observed in the absence of a vessel. This is explained by the heating of the hydrogel in the plane of the modeled vessel, which was the source of heat supply throughout the experiment. It is evident from Fig. 14 that the highest increase in temperature occurred in the first minute of thawing. Thus, the presence of the vessel model did not affect the values of the lowest surface temperature, although it did cause a «shift» of the coolest areas from the freezing zone center. Such results indicate that the ice lump played the role of a kind of cold accumulator, the «power» of which was sufficient to maintain the dynamics of the values of the lowest surface temperature in the «butterfly wings».

It should be noted that in our experiments, a urological catheter with proper thermal insulation was used as a vessel model (the catheter is made of polyvinyl chloride, the thermal conductivity of which is 0.16 W/mK). The choice of this catheter as a vessel model was dictated by our desire to maintain the flow of fluid even when the vessel model was frozen into ice. In addition, the liquid flowed through the vessel at a moderate speed and had room temperature. These limitations, laid down in the planning of the experiment, will be taken into account in our further experiments. Using a tube with greater thermal conductivity as a vessel model, as well as varying the speed and temperature of the liquid flowing through the vessel model, will allow us to obtain more information about the effect of heat influx from blood vessels on the shape, size and temperature fields of the frozen zone in real cryosurgical conditions.

**3.4. Comparison of experimental results with those of numerical simulation**

Under «ideal» cryoeffect conditions (point cryoapplicator, infinite volume of the object, no influence of extraneous factors, etc.), one can expect the formation of an ice lump shaped as a hemisphere and its further growth with a gradual slowdown in time and to the size determined by thermodynamic equilibrium conditions, depending on the ratio of the temperatures of the object and cold source. In this case, the temperature distribution has a radial nature and coincides on the surface as well as in the depth of the hemisphere, and then thermography is an ideal tool for remote monitoring of internal processes by the dynamics of surface thermal fields.

To verify the results obtained under the conditions of a «non-ideal» experiment, a mathematical model of thermal processes was developed under local exposure to low temperatures on the surface of a model object (5% gelatin hydrogel). The model describes the movement of the ice front and the dynamics of temperature distribution on the surface and inside the ice zone, including isotherms (0, -20 and -40) °C, during cryoexposure and natural warming. The model assumes the influence of external environment, the final surface area of the cryoapplicator, the case of partial immersion of the cryoapplicator in the hydrogel, additional cooling of the surface by means of convective air heat exchange with the cryoinstrument side walls [7].

The model uses a simplified approach to the Stefan problem (to describe the change in the phase state of a substance, in which the position of the phase boundary changes over time) using a non-uniform heat conduction equation with temperature-dependent parameters:

$$\frac{\partial T}{\partial t} = \frac{1}{\rho(T) Cp(T)} \nabla(k(T) \nabla T(r, z, t)) \qquad (3)$$

where T is the temperature, ρ is the density, $C_p$ is the heat capacity at constant pressure, k is the thermal conductivity, ∇ is the gradient of the scalar temperature field, r is the radial coordinate, z is the axial coordinate.

The thermodynamic task with such a large temperature difference and a moving boundary of phase change with complex geometry cannot be solved analytically. Therefore, to numerically solve the heat conductivity equation (3), the finite difference method was used on an isotropic cylindrical grid, each node of which has different temperature-dependent thermodynamic parameters (density, heat capacity, thermal conductivity) [7]. As an initial condition, we assumed that the entire sample has a uniform temperature $T_i$. For the boundary conditions, we assume that there is no heat transfer through the sides of the modeling region, and a constant temperature $T_i$ is maintained at the bottom. On the sides of the cryoapplicator, the temperature is kept constant at the temperature of liquid nitrogen, since the thermal adsorption of the cryoapplicator is significantly greater than the amount of heat supplied by the hydrogel and air. We include in our model the heat transfer through the air from the side walls of the cryoapplicator to the surface of the sample with the *effective thermal conductivity* [4, 13]. This allows to better describe two processes: fast decreasing of the radii of the 0°C and -20°C isotherms at the beginning of the thawing stage in Fig. 13 and also in Ref. [9], as well as fast increasing of the isotherms' radii during cooling. Note that this model can also be interpreted in terms of effective boundary conditions, which takes into account impact of the heat transfer, including the heat convection [12, 23].

In our simulation we use the one-node thickness in finite-difference method with the effective thermal conductivity $k_{eff}$ = 32 mW/(m*K) and effective heat capacity per unit volume $(\rho C_p)_{eff}$ = 560 J/(K*m$^3$), that is used to set up upper surface boundary conditions outside the cryoapplicator. Then we use these parameters to calculate the temperature spreading on the hydrogel-air border with Eqs. (23-29) from Ref. [7] for finite-difference method with non-uniform heat equation.

Using this formalism, we can calculate the spatiotemporal distribution of the temperature field near the cryoapplicator. The calculations were performed using the parameters of 5% gelatin hydrogel [1], including:

temperature-dependent density ρ:

$\rho = 935.79 + 8.715 \cdot 10^{-2} T + 1.7428 \cdot 10^{-4} T^2$, for $T \in [-160 \,°C, T_m]$;

$\rho = 1018.2 + 16.927 T + 1.2044 T^2 + 2.925 \cdot 10^{-2} T^3$, for $T \in [T_m, T_f]$;

$\rho = 935.79 + 8.715 \cdot 10^{-2} T + 1.7428 \cdot 10^{-4} T^2$, for $T \in [T_f, 40 \,°C]$;

temperature-dependent heat capacity $C_p$:

$C_p = 2.1 + 7.9 \cdot 10^{-3} T + 7.1 \cdot 10^{-6} T^2$, for $T \in [-160 \,°C, T_f]$;

$C_p = 3.95 + 1.25 \cdot 10^{-3} T$, for $T \in [T_f, 40°C]$;

temperature-dependent thermal conductivity k:

$k = 2.191 - 0.0118 \cdot T + 6.57 \cdot 10^{-5} T^2$, for $T \in [-160, -2.8] \,°C$;

$k = 0.615 + 0.00181 \cdot T$, for $T \in [-2.8, 40] \,°C$. (4)

Here $T_f$ = -0.1 °C is the cryoscopic temperature for gelatin hydrogel at which phase changing process starts during freezing, $T_m$=-14 °C is the initial melting temperature [1].

Figure 16 shows a schematic 3D image of the area of modeling thermal processes under local exposure to low temperatures on the surface of a model object (hydrogel) using a flat cryoapplicator of finite size, immersed 1 mm into the hydrogel. The figure shows the numerically calculated distribution of the temperature field after 1 minute of freezing. The upper plane demonstrates the temperature distribution on the hydrogel surface, which is consistent with thermographic observations of the surface thermal field at this moment of cryoexposure. The front vertical plane shows a section along the line $Y = 0$ (through the center of the applicator), which visualizes the deep thermal field. The blue, red and brown lines show the instantaneous shapes and locations of the surface and deep isotherms (0, -20 and -40) °C, respectively. Our calculations show that the shape of the ice ball in hydrogel is close to the semi-ellipsoid with focuses on the ridges of the cryoapplicator, at all moments of time. The respective half-ellipses are shown by thin dashed lines in Fig. 16. This is due to the insufficient insulation of the side walls, resulting in the cooling of the surface through the air leading to the formation of a thin disk of frozen hydrogel around the main ice spot on the surface. The calculation results presented in Fig. 16 show the relationship between the temperature distribution at the surface and in depth, which can be useful for predicting the freezing depth by analyzing the thermal imaging data from the surface. We also note that the dynamics of the low-temperature isotherms (for -20°C and -40°C) is not significantly influenced by taking heat transfer through the air, while for the zero-temperature isotherm this results in increasing the size of the circle on the surface.

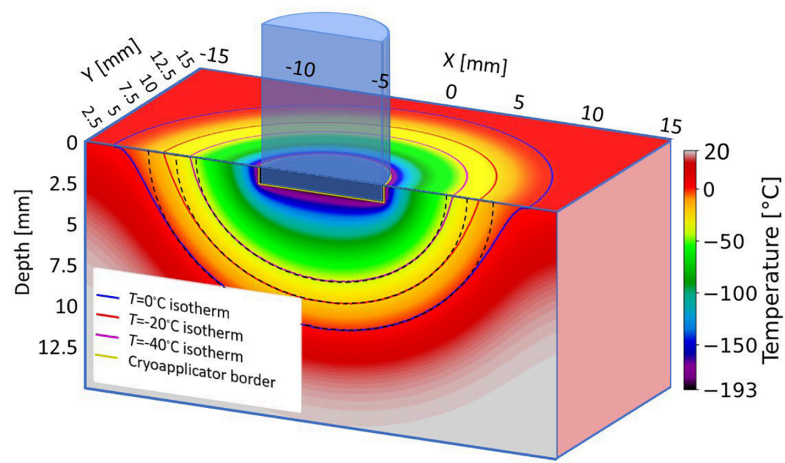

Fig. 16. 3D image of the simulation area of thermal processes under local exposure to low temperatures on the surface of a model object (hydrogel) by a flat cryoapplicator of finite size, immersed 1 mm into the hydrogel. The thermal fields on the surface (horizontal plane) and in depth (section in the frontal plane) correspond to the state after 1 min of cryoexposure. Blue, red and magenta lines show instantaneous shapes and locations of surface and depth isotherms (0,

−20 and -40) ◦C, respectively. Black dashed lines show instantaneous shapes and positions of surface and depth isotherms without heat transfer through air.

Figure 17 shows three calculated isotherms (solid lines) as they develop over the cryoexposure time. It should be noted that these calculations for solid lines take into account the experimentally observed fact of additional cooling of the hydrogel surface with cold air, and dashed line calculated without additional cooling of the hydrogel with cold air. For comparison, experimental data are shown as separate points (see Fig. 4). It is evident that the experimental points and curves agree well with each other. To conclude it should be noted that the agreement of these calculations with the measurement results (Fig. 17) demonstrates the usefulness of solving the heat equation, as described in detail in Ref. [7], that can be helpful for understanding the measurements, as well as predicting the three-dimensional thermal field distribution and dynamics.

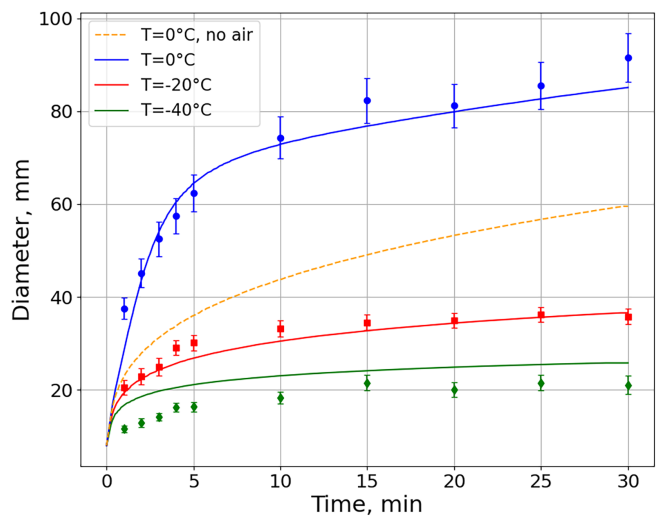

Fig. 17. Calculated increase in the diameters of surface isotherms (0, –20 and –40) °C during low-temperature exposure of a flat cryoapplicator to the hydrogel surface (solid lines) and 0°C without heat transfer through air (dashed line). Individual points are measured data. Mean (± S) of the diameters measured in five experiments.

In Fig. 17 we obtained quite good quantitative agreement between the experiment and theory for 0°C and -20°C, but for the -40°C curve we obtain only qualitatively good agreement, which might be caused by the fact that most of the extrapolations for the thermodynamic parameters in Eq. (4) were based on the experimental data down to -50°C [1]. It means that the thermodynamic parameters, for which the polynomial approximations were used, could differ from the real ones in the region between the isotherm -40°C and the cryoapplicator. So, the thermodynamic parameters in our Eq. (4) could differ from the real ones in the temperatures between -196°C and -50°C, i.e. closer to the cryoapplicator, and this leads to the increase of the thermal diffusivity coefficient, so the parameters, which we take, overestimate the value of the thermal diffusion coefficient. The isotherms 0°C and -20°C have a good agreement because they are located close to the phase change region between the temperatures $T_m$ and $T_f$, so the main impact on the temperature dynamics is determined by the latent energy of the phase transition, that is much larger than the heat capacity $C_p$ [7].

**Conclusions**

1. The developed measuring equipment enables the *in vitro* studies of freeze-thawing to predict the dynamics of freeze-thawing during cryodestruction of biological tissues *in vivo*. The equipment allows the varying of the conditions of low-temperature exposure and thawing, regulating the temperature and composition of the model fluid, simulating the presence of vessels with different blood flow parameters, etc. The use of this equipment allows simultaneous and synchronous assessment of the dynamics of the shape and size of the freezing zone, the temperature at specified points in the depth of the hydrogel and at any point on its surface during freezing and thawing.

2. When a flat cryoapplicator is applied to the hydrogel surface at low temperature, the one in its depth and on the surface rapidly decreases. A constantly growing ice lump is formed in the hydrogel, the base of which on the surface has the shape of a round ice spot. Surface isotherms with a temperature of (0, –20 and –40) °C are shaped as concentric circles. To verify and understand the physical aspects of these processes, we numerically solved the heat conductivity equation and obtained quantitative agreement between the calculated and experimental data on the dynamics of these isotherms during cooling, as demonstrated in Fig. 17.

3. When modeling a vessel with blood flow in the cryoeffect zone, the shape of the surface thermal fields is deformed, leading to the appearance of thermal fields in the form of butterfly wings. In this case, the surface isotherms of 0 °C and –20 °C undergo deformation from the first seconds of low-temperature exposure, and the –40 °C isotherm is deformed after 2 minutes of cryoexposure. There is also a shift of the most cooled areas from the center of the freezing zone to the area of the «butterfly wings».

4. During natural thawing, there is a slow decrease in the ice lump size and a gradual rise in temperature in its depth. The surface temperature of the hydrogel also increased, but at first sharply, which was caused by the cessation of «side» cooling of the surface by cold air. The dynamics of the decrease in the size of the ice lump with or without the blood vessel model are similar during 30 minutes of thawing that can be explained by the fact that the ice lump plays the role of a cold accumulator.

5. Undoubtedly, during cryoexposure with a quasi-point cryoapplicator, IRT plays an important role in monitoring the growth of the volumetric ice zone and assessing the dynamics of volumetric temperature fields. IRT is also a unique tool for identifying «unplanned» heat sources and analyzing their impact on the object of cryoeffect. However, the results of the analysis of surface thermal fields to deep thermal processes should be extrapolated with caution, taking into account the presence of secondary factors that affect the thermodynamics of the object under real conditions of cryoeffect. During short-term cryoexposure, the influence of these secondary factors is insignificant, whereas during long-term cooling, their influence can lead to significant errors in assessing the depth of freezing of the object based on the temperature fields on its surface. Taking into account the maximum duration of cryoexposure, at which the influence of secondary factors can be neglected, when developing cryosurgical protocols will allow obtaining reliable information on the shape, depth and distribution of temperatures in the frozen zone based on IRT data. Use of such devices will open new prospects to be applied in cryosurgical protocols in future.

**Funding**

This work was supported by the National Research Foundation of Ukraine (Grant No. 2022.01/0094).